\documentclass[referee]{raa}            

\usepackage{graphicx,times}             
\usepackage[figuresright]{rotating}
\usepackage{natbib}

\input{epsf.sty}                        
\input{psfig.sty}                       

\begin{document}

   \title{The Debris Disk Candidates: Eleven 24$\mu$m excess stars in {\it Spitzer} SWIRE Fields
}

   \volnopage{Vol.0 (200x) No.0, 000--000}      
   \setcounter{page}{1}          

   \author{Hong. Wu
      \inst{1}
   \and Chao-Jian Wu
      \inst{2}
   \and Chen Cao
      \inst{3,4}
\and Sebastian Wolf
      \inst{5}
 \and Jing-Yao Hu
      \inst{1}
   }

   \institute{Key Laboratory of Optical Astronomy, National Astronomical Observatories, Chinese Academy of
Sciences, Beijing 100012, P.R.\ China; {\it hwu@bao.ac.cn}\\
        \and
        Department of Astronomy, Beijing Normal University, Beijing, 100875, P.R.\ China\\
        \and
        Institute of Space Science and Physics, Shandong University at Weihai, Weihai, Handong 264209, P.R.\ China\\
        \and
        Shandong Provincial Key Laboratory of Optical Astronomy \& Solar-Terrestrial Environment,
Weihai, Shandong, 264209, P.R.\  China\\
    	\and
 Max Planck Institute for Astronomy, K\"{o}nigstuhl 17,
69117 Heidelberg, Germany\\
  }

   \date{Received~~2009 month day; accepted~~2009~~month day}

\abstract{ We present the optical to mid-infrared SEDs of 11 debris disk candidates from $Spitzer$
SWIRE fields.  All these candidates are selected from SWIRE 24$\mu$m sources matched with
both the SDSS star catalog and the 2MASS point source catalog. They show an excess in the
mid-infrared at 24$\mu$m ($K_S$-[24]$_{Vega}$ $\ge$ 0.44) indicating the presence of a
circumstellar dust disk. The observed optical spectra show that they are all late type
main-sequence stars covering the spectral types of FGKM.  Their fractional luminosities
are well above 5$\times$10$^{-5}$, even up to the high fractional luminosity of
1$\times$10$^{-3}$. The high galactic latitudes of SWIRE
fields indicate that most of these candidates could belong to the oldest stars in the thick
disk.  Our results indicate that the high fractional luminosity debris disks could exist in
the old solar-like star systems, though they are now still quite rare. Their discoveries at
high-galactic latitudes will also provide us an excellent opportunity to the further studies
of properties and evolution of the debris disk in the ISM poor environments.
}

   \authorrunning{Wu H. et al. }            
   \titlerunning{The Debris Disk Candidates in SWIRE Fields }  

   \maketitle

\section{Introduction}           
\label{sect:intro}

\par
One of the most notable achievement of the {\it InfraRed Astronomical Satellite} (IRAS) is
the discovery of dusty circumstellar disks \citep{zuckerman01}. An example is Vega, which
shows a large infrared excess from a main-sequence star \citep{aumann84}. This provided the
direct evidence to  the existence of a debris disk firstly. The scattered light from several
nearby debris disks has been analyzed by use of the HST \citep{krist05}. Debris disks are
thought to be formed from an dissipated  optically thick accretion disks left over from star
formation \citep{hollenbach00,wyatt02,gorlova04}. The study of debris disks is crucial to
understand the formation and existence of planets\citep{bryden09} and smaller objects, such
as comets and asteroids \citep{zuckerman01}.

Limited by the sensitivity of IRAS, most of the debris disk discovered were for nearby stars
much younger than the sun \citep{metchev04,bryden06}, since they are likely to be at a
transition stage with a higher frequency to possess a disk with a plenty of dust
\citep{zucker04,bryden06}, and their luminosities are high enough to heat a debris disk to
an IRAS-detectable level \citep{krist05}. With  the higer sensitivity, better resolution and
the more extended wavelength \citep{jourdain99}, the {\it Infrared Space Observatory (ISO)}
has expanded our capability of searching for circumstellar dust \citep{Spangler01} to the
modest infrared excess among the older stars \citep{decin00,decin03,habing01}.

With the launch of the {\it Spitzer} Space Telescope \citep{werner04}, a new level of
sensitivity and spatial resolution is provided to the studies of debris disks in the infrared
\citep{su05}. It provides the potential to identify and investigate the debris systems that
were not detectable with previous observatories \citep{kim05}. It also extends the search for
debris disks to larger distances \citep{bryden06}. Several {\it Spitzer} programs have focused
on this field, including the Cores to Disks \citep[C2D]{evans03}, Formation and Evolution
of Planetary Systems \citep[FEPS]{meyer04}, the Galactic Legacy Infrared Mid-plane Survey
Extraordinaire  \citep[GLIMPSE]{benjamin03}, and the Survey of Solar-Type Stars
\citep[SSS]{beichman05}. However, most of these programs are designed for the search of
the infrared excess in either nearby known planetary systems or the debris systems at regions,
where interstellar material is rich, such as the low galactic latitude regions and the molecular
clouds. While, at high Galactic latitudes, there are also some wide-area surveys with the
Multiband Imaging Photometer for {\it Spitzer} \citep[MIPS]{rieke04}, such as the NOAO Deep
Wide-Field Survey  \citep{jannuzi99,houck05,brand06}, the {\it Spitzer} Wide-area Infrared
Extragalactic Survey  \citep[SWIRE]{lonsdale03}, and the {\it Spitzer} extragalactic First Look Survey
 \citep[xFLS]{fadda06}, which provide us more star candidates with infrared excesses
\citep{hovhannisyan09}.

The {\it Spitzer} SWIRE were observed in the four mid-infrared IRAC bands (3.6, 4.5, 5.8, 8.0~$\mu$m)
\citep{fazio04} and the three mid-to-far infrared MIPS bands (24, 70, 160~$\mu$m) \citep{rieke04}.
Being designed for extragalactic surveys, all the six SWIRE fields are at high galactic latitudes
and cover about 50 sq. degree. The wide field and high sensitivity provide us an opportunity to
search for the new faint debris disk candidates even the planets\citep{bryden09} in the thick disk or
even in the halo. Contrary
to the other $Spitzer$ programs mentioned above, which focus on the younger or nearby systems,
the debris disk candidates selected from SWIRE fields are much farther in distance from 300 to
2,000 pc compared with those which have a distance from 10 to 150 pc in FEPS, and also much
older in age of about 10 Gyr belonging to the oldest systems. Generally, the star that is older than
10 Myr can be regarded as a candidate with debris disk\citep{gorlova06}. This sample will help us to
explore the formation of planetary systems in ISM poor environments.

In this paper, we first describe the optical to infrared observations, data reduction and
candidate selection in Section 2.  In Section 3, we give the infrared properties of these
24$\mu$m excess stars, their spectral energy distributions (SEDs). The discussions
and summary are presented in Section 4 and 5.

\section{DATA, OBSERVATIONS AND CANDIDATE SELECTION}
\label{sect:Obs}
\subsection{{\it Spitzer} Mid-Infrared Data and Data Reductions}

The 24$\mu$m excess stars are from ELAIS-N1, ELAIS-N2, Lockman Hole fields of the SWIRE fields.
The common regions of observations between the IRAC and MIPS for ELAIS-N1, ELAIS-N2, and
Lockman Hole fields are about 8, 4, and 11 sq. degree \citep{surace05}. The BCD (Basic Calibrated
Data) images of the IRAC four bands were  obtained from {\it Spitzer} Sciences Center (SSC), which
include flat-field corrections, dark subtraction, linearity and flux calibrations \citep{fazio04}.
The IRAC images were mosaiced from the BCD images after pointing refinement, distortion correction
and cosmic-ray removal with the final pixel scale of 0.6\arcsec \ as described by \citet{huang04}
and \citet{wu05}; Whilst the MIPS 24$\mu$m images were mosaiced in the similar way with the final
pixel scale of 1.225\arcsec \citep{wen07,cao07}.  Matching the sources detected by SExtractor
\citep{bertin96} in the five bands (IRAC four bands and MIPS 24$\mu$m band ) with the 2MASS
sources, we obtained the astrometric uncertainties of less than 0.5\arcsec.

The mid-infrared photometries were obtained by SExtractor with an aperture of 3\arcsec \ for the
IRAC four bands and 10.0\arcsec \ for the MIPS 24$\mu$m band. All these magnitudes are in AB
magnitude system \citep{oke83}.
All the magnitudes of four IRAC bands were corrected  the aperture of 24\arcsec.
Comparing with the model colors of IRAC magnitudes and 2MASS $K_S$ magnitude \citep{cutri03} for
the 2MASS stars with $J$-$K_S$ $\le$ 0.3 as \citep{lacy05} did for sources in {\it Spitzer}
xFLS field, the small additional corrections were made for all
the four IRAC bands. And these give us the calibration errors in the four IRAC bands better than
0.08 mag. As for 24$\mu$m, the magnitudes were first corrected to the aperture of 30\arcsec.
Then an additional  correction was also performed according to \citet{surace05}.
The calibration accuracy of 24$\mu$m is better than 10\%
\citep{rieke04}. The 5$\sigma$ flux in the IRAC four bands and MIPS 24$\mu$m band are about
5, 8, 43, 43 and 200 $\mu$Jy respectively.

\subsection{Optical and Near-IR Photometries}

The ELAIS-N1, Lockman Hole and ELAIS-N2 fields are completely or partly covered by the Sloan
Digital Sky Survey \citep[SDSS]{stoughton02} and the total overlap regions used in this work
is about 15 sq. degree. All the $Spitzer$ sources were matched with the SDSS point sources from
"Star" catalog of Data Release 4 \citep{adelman06} with a radius of 3.0 $\arcsec$. The PSF
magnitudes of $u$,$g$,$r$,$i$,$z$ bands were adopted in this work. The near-infrared magnitudes
were from the 2MASS point source catalog \citep{cutri03}. The default $J$, $H$, $K_S$ band
magnitudes were adopted.

\subsection{Candidate Selection}

\citet{gorlova04} studied the 24$\mu$m emission of the M47 stars, and found that the majority
of them have a mean $K_S$-[24]$_{Vega}$ value of 0.11 with a $\sigma$ of 0.11 if described by a
Gaussian distribution. Therefore they defined the 24$\mu$m excess stars as those 3$\sigma$ redder
than the mean value, that is, $K_S$-[24]$_{Vega}$ $\ge$ 0.44.  We adopted their selection
criterion to select 24$\mu$m excess stars.
[24]$_{Vega}$=[24]$_{AB}$-6.74
was used to transform the AB magnitudes in the 24$\mu$m band to the Vega magnitudes. To obtain
the optical spectra, we only selected stars with $r$ magnitudes less than 17.5. We also removed
the objects with fuzzy features, whose 24$\mu$m emission could come from the background galaxies.
Finally, we selected 11 24$\mu$m excess candidates. Figure 5 shows the locations of all the
candidates in the color-color diagram and the selection criterion is also plotted as a vertical
dotted line. Their names and the optical to infrared magnitudes of the candidates are listed
in Table 1. Since there are problems in the $r$ magnitude of the star J163754.26+405259.1 and
the $g$,$r$,$i$ magnitudes of the star J163948.68+413711.0, we did not include these measurement
in our subsequence analysis. From Table 1, we can see that except J104508.69+592830.5, all these
candidates have $K_S$-[24]$_{Vega}$ value above 1, indicating strong 24$\mu$m excess. The optical
$r$-band, IRAC 3.6$\mu$m, 8.0$\mu$m and MIPS 24$\mu$m images are shown in Figure 1. The central
circle in each image indicates the optical position of each star with a radius of 1.5\arcsec.
It can be seen that all positions in the four bands are consistent.

\subsection{Optical Spectroscopy}

The optical spectra of 11 debris disk candidates stars were obtained with the 2.16m telescope
at Xinglong, NAOC (National Astronomical Observatories, Chinese Academy of Sciences)  from
February to May, 2006. The attached spectrograph is either the OMR spectrograph with dispersion
of 200\AA/mm or BFOSC (Beijing Faint Object Spectrograph and Camera) with the grism G4. Both
spectrographs give a resolution of $\sim$ 10\AA\  and cover the wavelength range from 3800\AA\
to $\sim$ 8000\AA. The exposure times depend on the apparent magnitudes of these stars from 900
to 3600 seconds. The detail informations of the spectral observations are presented in Table 2.
All these spectral data were reduced by the standard procedures with IRAF packages, which include
overscan correction for BFOSC only, bias subtraction, flat-field correction. The He/Ne/Ar and
Fe/Ar lamps were used for the wavelength calibrations of the OMR and BFOSC spectra. The KPNO
standard stars were obtained to perform the flux calibration at each night. All resulting spectra
are shown in Figure 2. The spectral classifications are given in Table 2. For those with a poor
signal-to-noise spectra (Stars 3, 8, 11), we had to determine their spectral types according
their SEDs. All these 11 stars are the main-sequence dwarf stars  with the spectral types from
M to F, and belong to solar-like  stars.

\subsection{Reddening}

Since all SWIRE regions are  at high galactic latitudes, the extinction by the Milky Way
extinction $A_V$ of all these stars are no larger than 0.04 from the SDSS "stars" catalog
\citep{adelman06} and are therefore neglected in the further analysis. How about the
local extinction of these stars?  Figure 3 shows the optical color-color diagram for
9 stars. Due to lack of some optical bands, the stars J163948.68+413711.0 and J163754.26+405259.1
are not plotted in this figure. As a comparison, the models of main-sequence stars and giants \citep{fitzgerald70} are
also presented as the different curves.
All 9 stars are consistent with late type stars as above and are located nearby the curve of the
model stars. This indicates that all these stars are slightly obscured.
The extinction of $K_S$ is only 10\% of $A_V$ and that of 24$\mu$m is even smaller \citep{ schlegel98, mccall04, rieke05}.
Therefore all extinctions can be neglected.

\section{RESULTS AND ANALYSIS}

\subsection{IRAC Colors}

Figure 4 shows the mid-IR color-color plot for the 11 stars in IRAC bands. The small circles
represent the positions of the stars and their corresponding numbers are also labelled. Vega is plotted as a star symbol. There are not any obvious deviation of colors (either [3.6]-[4.5] or [3.6]-[8.0]) for Stars 1,4,6,9 from Vega. As Vega has proved to be an IR excess star,
for comparison we also plot the position of black-body with Vega temperature of 9600K as plus symbol. Both those stars and Vega show minor excess from black-body in [8.0] band. However, Stars 5,7,8 show large deviation (more than 1 magnitude) in color [3.6]-[8.0]. Such a 8$\mu$m excess can be seen obviously in their SEDs in Figure 6. The remaining Stars 2,3,10,11 show a marginal or modest deviation in [3.6]-[8.0], which show small excess at 8$\mu$m in SEDs (Figure 6). Only two (Stars 3,7) of these stars show a marginal deviation in [3.6]-[4.5], and the deviations are about 0.2, about 2-3 times typical error. Therefore, we can be sure that most of these 24$\mu$m excess stars show an excess in the 8$\mu$m band, and few of them show an obvious excess in the 4.5$\mu$m band.

\subsection{MIPS 24$\mu$m Excess and Fractional Luminosity}

Figure 5 presents the $J$-$H$ versus $K_S$-[24]$_{Vega}$ diagram. The dashed are the
curve of main sequence stars from \citet{gorlova04}. All the 11 stars are located
at the upper late type stars region. This is consistent with the result of the spectral
classification.  Almost all of these stars present a large 24$\mu$m excess. Their
$K_S$-[24]$_{Vega}$ colors are larger than 1, even to 6. These values are far above our
selection criterion value of 0.44.  Only the late type Star 9 has a $K_S$-[24]$_{Vega}$ value
of 0.51, little larger than 0.44.

Unfortunately, we can not obtain the confirmed far-infrared fluxes for all these stars, because
of the low resolution and low sensitivity of the $Spitzer$ MIPS 70$\mu$m and 160$\mu$m bands.
Without the total infrared luminosities of stars, we can not obtain the fractional luminosity $f_d$,
which is defined as the ratio of integrated infrared excess of the disk to bolometric luminosity
of star \citet{moor06}, to characterize  the amount of dust. To roughly estimate the fractional
dust luminosities, we assumed $F_{IR}\sim \nu F_{\nu}[24\mu m]$, as \citet{chen05a}. The
bolometric luminosities of stars are from \citet{jager87}. The calculated fractional luminosities
are listed in Table 2. The fractional luminosities range from  5$\times$10$^{-5}$ to
3$\times$10$^{-3}$. It is quite interesting that four of those stars even have the fractional
luminosity values higher than 10$^{-3}$.

\subsection{Spectral Energy Distributions}

The SEDs of 11 stars are shown in Figure 6. They cover the wavelength range from the optical to
the mid-IR bands, including the available photometries of SDSS $u$,$g$,$r$,$i$,$z$,
2MASS $J$,$H$,$K_S$, $Spitzer$ IRAC four bands and MIPS 24$\mu$m band. These stars exhibit a
variety of mid-IR properties. Five of them show excess only at 24 $\mu$m and four show obviously
excess at both 8$\mu$m and 24$\mu$m. Star 3 even presents excess at shorter wavelength of
3.6$\mu$m.  Except Star 9, all the stars show an deviation from the photosphere continuum at
mid-IR, indicating the existence of an inner hole of the dust disk \citep{muzerolle06,young04}.

\subsection{Notes For Individual Source}

\noindent {\bf J160551.07+534841.0, J160650.59+543420.6, J163754.26+405259.1, J163948.68+413711.0,
J104205.94+594657.2:} These stars have high S/N spectra and a moderate 24$\mu$m excess (Fig 5) with
$K_S$-[24]$_{Vega}$ between 1 and 4. Their SEDs from the optical to IRAC bands are well consistent
with photosphere continuum and only present infrared excess at 24$\mu$m (Fig 6).

\noindent {\bf J104508.69+592830.5:} This is a M-type star from its spectra (Fig 3). Though it satisfies
the
$K_S$-[24]$_{Vega}$ criterion, it presents a marginal excess above the photospheric emission at
24$\mu$m.

\noindent {\bf J163236.05+405537.3:} This is a G-type star and the only source with 8$\mu$m flux higher
than 24$\mu$m flux in the sample (Fig 3 and Fig 6).  Such a feature indicates the peak of infrared emission just between
8$\mu$m and 24$\mu$m and the outer radius of the dust disk would be much smaller than 100 au.

\noindent {\bf J104537.18+570532.9:} Though it has a low S/N spectra, combining with its optical and
near-infrared colors, we still can classify it as a K-type star. Its SED shows that there is little
deviation from photospheric emission at 8$\mu$m. However, it presents a high infrared excess at
24$\mu$m ($K_S$-[24]$_{Vega}$$\sim$5).

\noindent {\bf J163611.64+412427.9, J163730.40+403553.1:} As above, with the low S/N spectra,
J163730.40+403553.1 was classified as a K-type star with the help of either the colors and the SED.
J163611.64+412427.9 is a G-type star. Both stars show a high infrared excess at 24$\mu$m.
Their SEDs show an abrupt increase from 5.8$\mu$m  to 8$\mu$m and then
continue to increase to 24$\mu$m.

\noindent {\bf J160122.04+545708.2:} For this source, we only have very low signal-to-noise spectra
and classified it as a K-type star. With $K_S$-[24]$_{Vega}$ of 5.62, it is the strongest 24$\mu$m
excess source among the 11 stars. Its SED shows a deviation from photospheric emission even in the
near-infrared band, indicating its smaller inner radius of dust disk.

\section{DISCUSSION}

\subsection{The Coincidence Probability}

Though we excluded the sources with the fuzzy features, it is still possible that some background
and distant galaxies coincide with the position of the candidates and contribute to the measured
radiation at 24$\mu$m. Therefore, we need to estimate such a coincidence probability for each source.
As \citet{stauffer05}, we first obtained the cumulative source counts for those with 24$\mu$m magnitude
brighter than the star itself. Here we give an example to explain how we estimated the coincidence
probability of the star J160551.07+534841.0. Since it has 24$\mu$m magnitude of 16.1, we can estimate
the cumulative sources counts of 7.8$\times$10$^{5}$ per steradian for sources with the 24$\mu$m
magnitude less than 16.1 in the SWIRE ELAIS-N1 fields based on our catalog. This is consistent with
that of order of 7$\times$10$^{5}$ per steradian for flux density greater than 1.3 mJy
(16.1 magnitude in AB system) \citep{papovich04} from the another high galactic latitude field
$Spitzer$ xFLS. It corresponds to about one source per 54,000 arcsec$^2$, considering the previous
matched radius of 3 arcsec. The coincidence probability of the background 24$\mu$m source with
magnitude less than 16.1 being close to the line of sight to J160551.07+534841.0 is about 0.0005.
Similarly, we  estimated  the  coincidence probabilities of the 11 stars are among 0.0003 and 0.003.
Since all the stars locate at the high galactic latitude, it is impossible to be contaminated by cirrus.
However, we still can not exclude the possible contamination by background AGNs \citep{stauffer05}.

\subsection{Comparison with Other 24$\mu$m Excess Samples}

Until now, most detected 24$\mu$m excess stars observed by $Spitzer$ are young systems. \citet{low05}
presented 24$\mu$m observation for 24 members of the 8-10 Myr old TW Hya association. They found four
24$\mu$m excess stars. \citet{young04} detected several 24$\mu$m excess stars from young (25 Myr)
cluster NGC~2547. Almost all of these stars have $K_S$-[24]$_{Vega}$ colors less than 2. Only one
shows a large $K_S$-[24]$_{Vega}$ color up to 4. \citet{gorlova04} selected 24$\mu$m excess stars
from 100 Myr old open cluster M~47. Seven of the early type stars show smaller excess with
$K_S$-[24]$_{Vega}$=0.6-0.9 and one early type has a color value of 2.4. Three late type stars have
$K_S$-[24]$_{Vega}$ colors between 1 to 4. By studying the 100 Myr Pleiades cluster, \citet{gorlova06}
obtained an excess fraction of 25\% (5/20) for  the early-type stars and 10\% (4/40) for solar-type
stars. All of these stars have $K_S$-[24]$_{Vega}$ colors less than 1.5.  \citet{chen05a} obtained
the observations of 40 F- and G-type members of the Scorpius-Centaurus OB association with ages
between 5 and 20 Myr at 24$\mu$m. They detected 14 objects that posses 24$\mu$m fluxes $\ge$ 30\%
larger than their predicted photosphere. \citet{chen05b} obtained the observations of 39 A-through
M-type dwarfs with estimated ages between 12 and 600 Myr. Only three stars possess a 24$\mu$m excess.
\citet{su06} reported the 24$\mu$m measurements of ~160 A-type main-sequence stars with ages ranging
from 5 to 850 Myr. The 24$\mu$m excess rate is 32\%. \citet{beichman05} searched for the infrared
excess debris disks toward 26 FGK field stars known to have one or more planets. All these stars
have a median age of 4 Gyr. None of them show an excesses at 24$\mu$m. Including the
\citet{beichman05}'s sample, \citet{bryden06} extended a well-defined sample of 69 FGK main-sequence
field stars also with the median age of 4 Gyr. Only one star shows excess emission at 24$\mu$m.
\citet{beichman06} searched for the circumstellar dust around a sample of 88 F-M stars, all detected
high S/N 24$\mu$m emission, but only a few present weak 24$\mu$m excess, though 12 of them present
a significant excess 70 $\mu$m emission.  \citet{morales06} identified two new debris disk candidates
with 24$\mu$m excess at high galactic latitude SWIRE fields. \citet{fajardo-acosta04} also searched
for the 24$\mu$m excess of main-sequence stars in another high latitude xFLS field.  \citet{koerner10}
presented 49 debris disk candidates that were within 25 pc of the Sun and with $V<9$. The candidate
24$\mu$m stars we found in the SWIRE fields are the solar-like stars with possible the oldest ages
(see the following section). Most of them present large 24$\mu$m excess with $K_S$-[24]$_{Vega}$ even
approaching 6.  Nowadays, only a few main-sequence stars have comparable large 24$\mu$m excess.

\subsection{The Age}

Though we have obtained the optical spectra of these 24$\mu$m excess candidates and classified them
based on their spectra and SEDs, we can not determine their ages based on any of the five methods of
age estimation for the main-sequence stars \citep{lachaume99}.  The low resolution of our spectra, and
even poor S/N for some faint stars, prevented us to obtain the ages by either the rotation or the iron
abundance methods, which need high spectral resolution to measure the rotational velocities or fine
absorption lines of different metal elements. The absence of calcium emission and the large uncertainty
of the distance determination of all these stars make it impossible to obtain the ages based on calcium
emission lines and isochrones. Since all these stars are the field stars and not born in a molecular
clouds, we can not use the kinematics method, too.

So what we can do is to determine their possible location (disk or halo) in our galaxy. As we know now,
there are three components except the bulge in our galaxy. They are the thin disk, the thick disk and
the halo. The previous works \citep{bahcall84,gilmore84,ojha99,chen01,du03} showed that the scale height
of the thin disk were in the range of 240 to 330 pc and the scale height of the thick disk is in the
range of 580 to 1300 pc. Recently, \citet{du06} analyzed 21 BATC fields with 15 intermediate-band filters.
The scale height they obtained for the thin disk varies from 220 to 320 pc and those for the thick disk
varies from 600 to 1100 pc, which is consistent with previous results. Considering that all the three
SWIRE fields are at high galactic latitudes of about 40 degree to 50 degree, more than half of the
candidate stars have a vertical distance to galactic plane above 600 pc (estimated from their spectral types), which comparable to the scale
height of the thick disk. Therefore, these stars could belong to the thick disk or even the halo.
The thick disk is an old component with an age around 10 Gyr \citep{gilmore95,gilmore89}.  It is quite
possible that many of these candidates belong to the population of the very old stars.

\subsection{The Fractional Luminosity}

There are two major zones of debris in the solar system: the asteroid belt at 2$-$4 au composed of
rocky material that is ground up by collisions to produce most of the zodiacal dust cloud and the Kuiper
Belt (KB) that consists of small bodies orbiting beyond Neptune's orbit at 30 $-$ 50 au \citep{kim05}.
The excess emission at the mid-IR band {\it Spitzer} 24$\mu$m and {\it IRAS} 25$\mu$m are sensitive to
material at several au \citep{su06}. Therefore, the stars with 24 $\mu$m excess are probably those with
an planetary systems containing a planetesimal belt corresponding to the asteroidal zone of the solar
system \citep{gorlova06}. From combination of observation and modelling, \citet{backman93} and
\citet{dermott02} determined that the fractional luminosity of our asteroid belt is $10^{-8}$ to $10^{-7}$.

\citet{artym96} argued that debris disks are confined to $f_d < 10^{-2}$ and sources with higher
fractional luminosity probably contain a significant amount of gas (e.g. T Tau and Herbig Ae/Be stars,
"transition" objects). \citet{zucker04} hypothesized that the stars with $f_d > 10^{-3}$ are younger
than 100\,Myr, and therefore a high $f_d$ value can also be used as an age indicator. This is supported by
Observations with the $ISO$  by \citet{beichman05}, who suggested a decline in the fraction of stars
with the excess IR emission with time, but still with the possibility of existing modest ($f_d >10^{-5}$)
excesses among the older stars \citep{decin00, decin03}. Meanwhile, \citet{bryden06} selected stars
without regard to their age, metallicity, or any previous detection of IR excess. The stars have a median
age of $\sim$4 Gyr. Their result shows that the debris disks with $f_d \geq 10^{-3}$ are rare around old
FGK stars.

On the contrary, \citet{decin03} claimed the existence of high $f_d$  disks  around older stars. Based
on the general evolutionary trend described by \citet{moor06}, the fractional luminosities of individual
systems were found to show a large spread of $10^{-6}< f_d < 10^{-3}$ at almost any age
\citep{decin03}. Particularly interesting is the relatively high number of older systems (t$>$500\,Myr)
with the high values of the fractional luminosity ($f_d \simeq 10^{-3}$).

Most of our candidates have the values of the fractional luminosity between 10$^{-5}$ to 10$^{-4}$ and
four of them even present the high fractional luminosities greater than $10^{-3}$. All the four stars
have a vertical distance to the disk above 900 pc and have the high possibility to be the old solar-like
stars in either the thick disk or in the halo. Therefore, they are possibly the rare systems up to date.

\subsection{Physical Explanation and Challenge}

How such a large amount of dust still exists in very old solar-like stars? Due to the effect of radiation
pressure, Poynting-Robertson drag and collisional destruction, the lifetime of the dust grains are quite
short \citep{beichman05,rieke05,backman93,lagrange00,dominik03}, for example, the dust grain with the
smaller size ($\leq$ 1$\mu$m) has a blown out time less than 100 yrs and the larger dust grain drifted
by Poynting-Robertson drag would be destroyed on the timescale of 1-10 Myr \citep{kim05}. These are all
far shorter than the age of the central stars. Therefore the dust observed must have been recently
produced \citep{beichman05}. \citet{rieke05} pointed out that these reproduced  dust in the "debris disk"
would arise primarily from collisions between planetesimals and from cometary activity. However, the
current model of gas planet formation and isotopic evidence from terrestrial and lunar samples indicate
that Jupiter mass planets and Earth-Moon system would  formed within several tens millions of years
\citep{silverstone06, kleine02,kleine03}. This still can not explain the high fractional luminosity
phenomenon of the old stars. They raise a serious challenge even to the new generation of theoretical
models \citep{moor06}. The most likely explanation for the presence of debris disk with the high
fractional luminosity at ages well above Gyr is delay onset of collisional cascades by late planet
formation further away from the star \citep{dominik03}. However whether such a mechanism can also
explain the very old star with an age of 10 Gyr is still questionable.

Besides, the most important is to  confirm the ages of these stars with the high fractional luminosity.
Therefore the higher S/N and higher resolution spectra are needed in the future observations.

\section{SUMMARY}

We present 11 debris disk candidates with distance from 300 to 2000pc and age about 10Gyr from
$Spitzer$ SWIRE fields, with the help of SDSS star catalog and 2MASS point source catalog. They
all present 24$\mu$m excess. The optical spectra obtained from 2.16 telescope at Xinglong, NAOC
and SEDs from the optical to mid-infrared are also presented. All these observations shows that:

All of these stars present the late type spectra and the infrared fractional luminosities from
5$\times$10$^{-5}$ to 3$\times$10$^{-3}$. They would be solar-like stars with the planetary debris disks.
Except J104508.69+592830.5, the SEDs of all other stars indicate the existence of an inner hole of the
dust disk.

We infer that many of these candidates would have ages of 10 Gyrs, since they are located in the thick
disk or in the halo of our galaxy.  Therefore, four of these stars could belong to the oldest stars with
the high fractional luminosities ($f_d$ $\ge$ 1$\times$10$^{-3}$).  Though they are quite rare until now,
we indicate that the high fractional luminosity debris disks can exist in the old solar-like star systems.

With the release of {\it Wide-field Infrared Survey Explorer} (WISE) data, there will be
more and more candidates to be found \citep{wright10}. The discoveries of the debris candidates at the high galactic
latitudes will provide us an opportunity to study the properties and evolution of the debris disk evolution
in ISM poor environments.

\begin{acknowledgements}

The authors would like to thank Jia-sheng Huang, Zhong Wang, Jia-rong Shi and Jing-kun Zhao for useful discussion and advice in $Spitzer$ data reductions and spectral classification. Also thanks referee's insightful comments. This project is supported by NSFC grants 11173030, 11078017, 10833006, 10978014, 10773014, and  partly supported by  China Ministry of Science and Technology under State Key Development Program for Basic Research (2007CB815400 and 2012CB821800). S. Wolf was supported by the German Research Foundation (DFG) through the Emmy Noether grant WO 857/2;.

This work was supported by the Key Laboratory of Optical Astronomy,
National Astronomical Observatories, Chinese Academy of Sciences.
This Also this work, in part, based on observations made with the Spitzer Space Telescope, which is operated by Jet Propulsion Laboratory of the California Institute of Technology under NASA Contract 1407.
Funding for the SDSS and SDSS-II has been provided by the Alfred P. Sloan Foundation, the Participating
Institutions, the National Science Foundation, the U.S. Department of Energy, the National Aeronautics
and Space Administration, the Japanese Monbukagakusho, the Max Planck Society, and the Higher Education
Funding Council for England. The SDSS Web Site is http://www.sdss.org/.
This publication makes use of data products from the Two Micron All Sky Survey, which is a joint project
of the University of Massachusetts and the Infrared Processing and Analysis Center/California Institute
of Technology, funded by the National Aeronautics and Space Administration and the National Science Foundation.

{\it Facilities:} Spitzer, Sloan, 2.16m telescope (NAOC).
\end{acknowledgements}

\clearpage
\begin{figure*}
\centering
\includegraphics[width=\textwidth, angle=0]{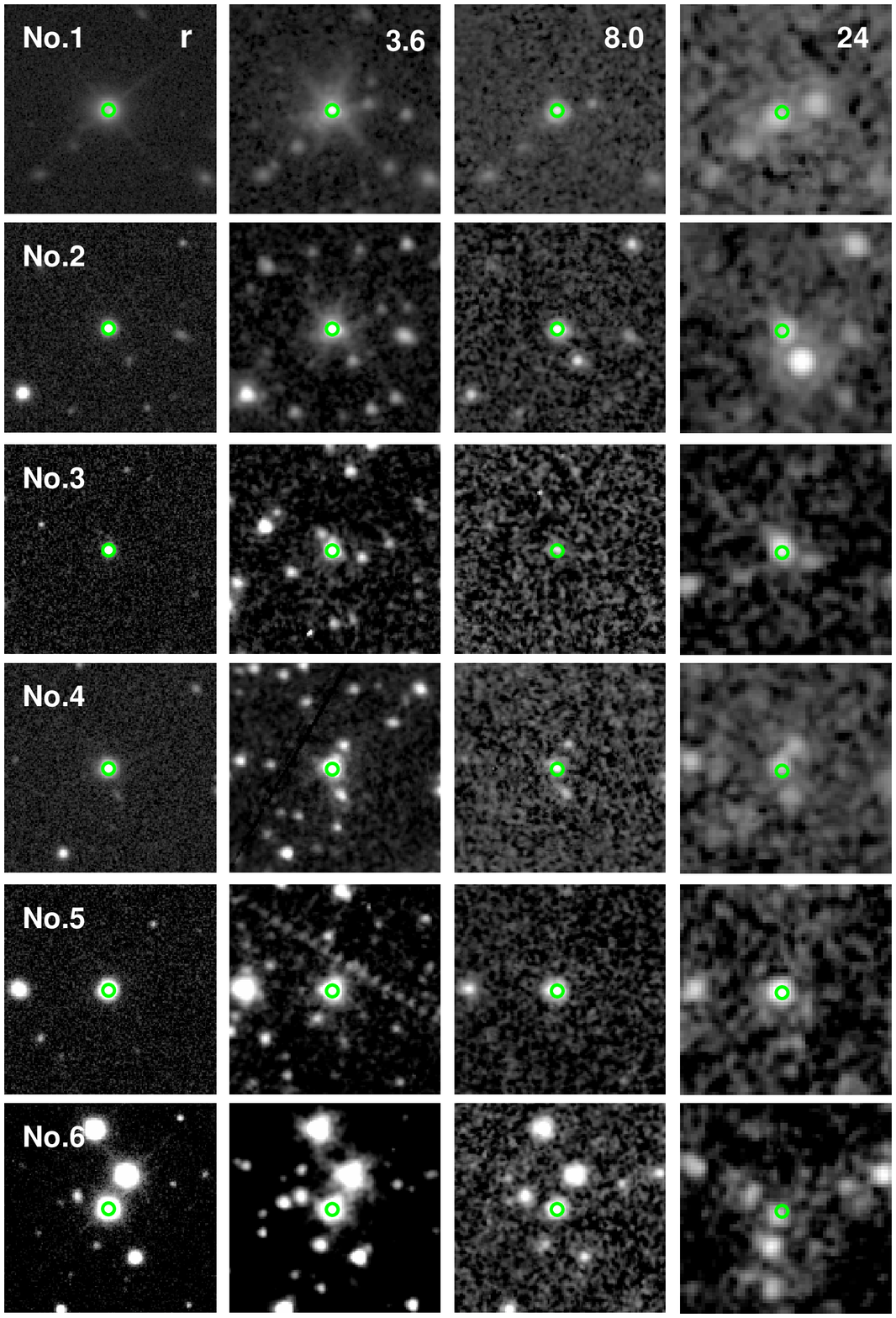}
\caption{ The SDSS $r$-band, Spitzer 3.6, 8.0 and 24$\mu$m bands images of the 24$\mu$m excess stars.
The circle in each image gives the position of stars in $r$-band with a radius of 1.5".\label{fig1}}
\end{figure*}

\clearpage
\begin{figure}[htbp]
\centering
\renewcommand{\thefigure}{1}
\includegraphics[width=\textwidth, angle=0]{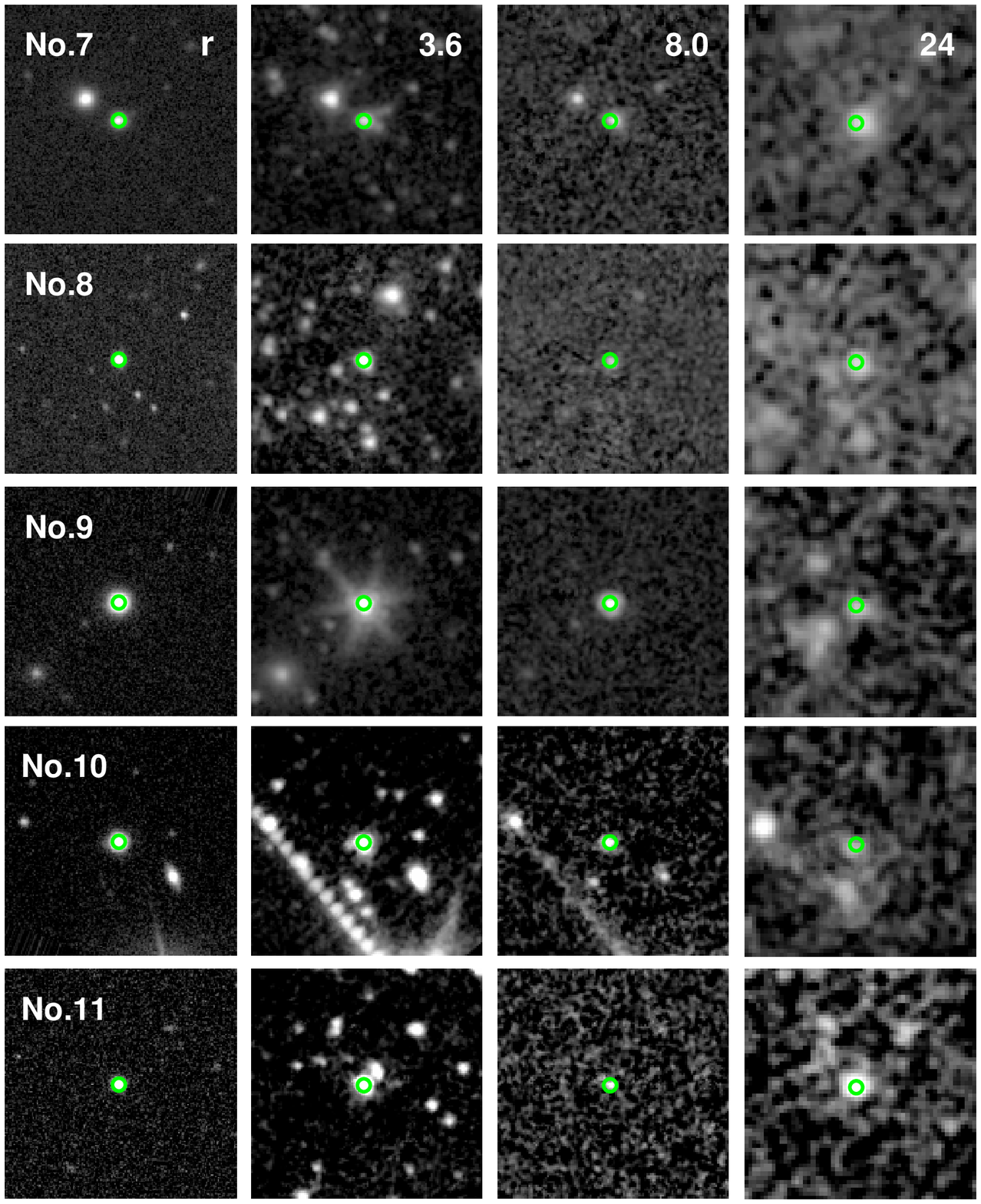}
\caption{Continued.\label{fig1}}
\end{figure}

\clearpage
\begin{figure}
\centering
\renewcommand{\thefigure}{2}
\includegraphics[width=\textwidth, angle=0]{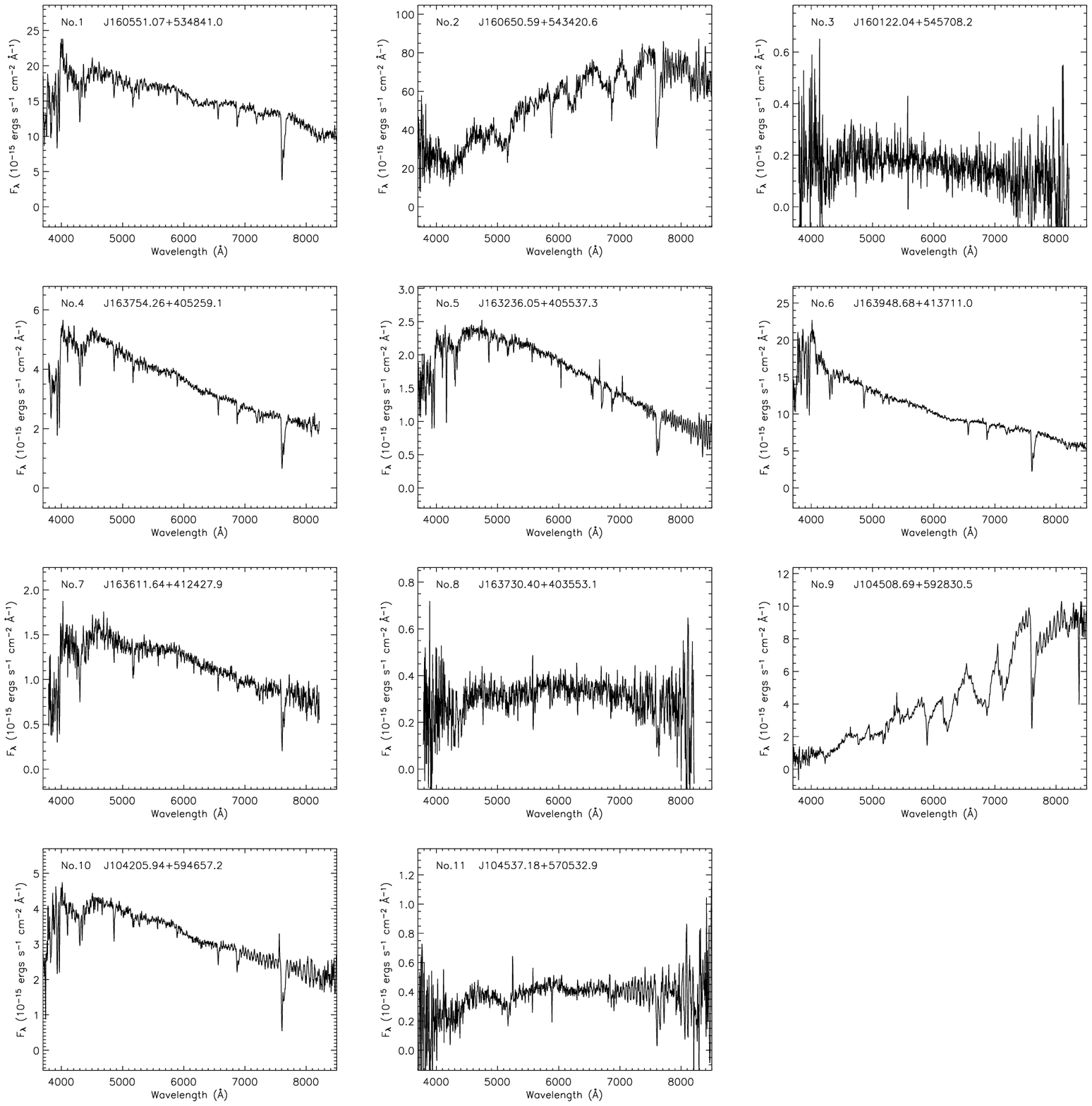}
\caption{The observed low resolution optical spectra of the 24$\mu$m excess stars. Star 3, 8 and
11 have very low S/N. All the stars present the late type star features.}
\label{fig2}
\end{figure}

\clearpage
\begin{figure}
\centering
\renewcommand{\thefigure}{3}
\includegraphics[width=\textwidth, angle=0]{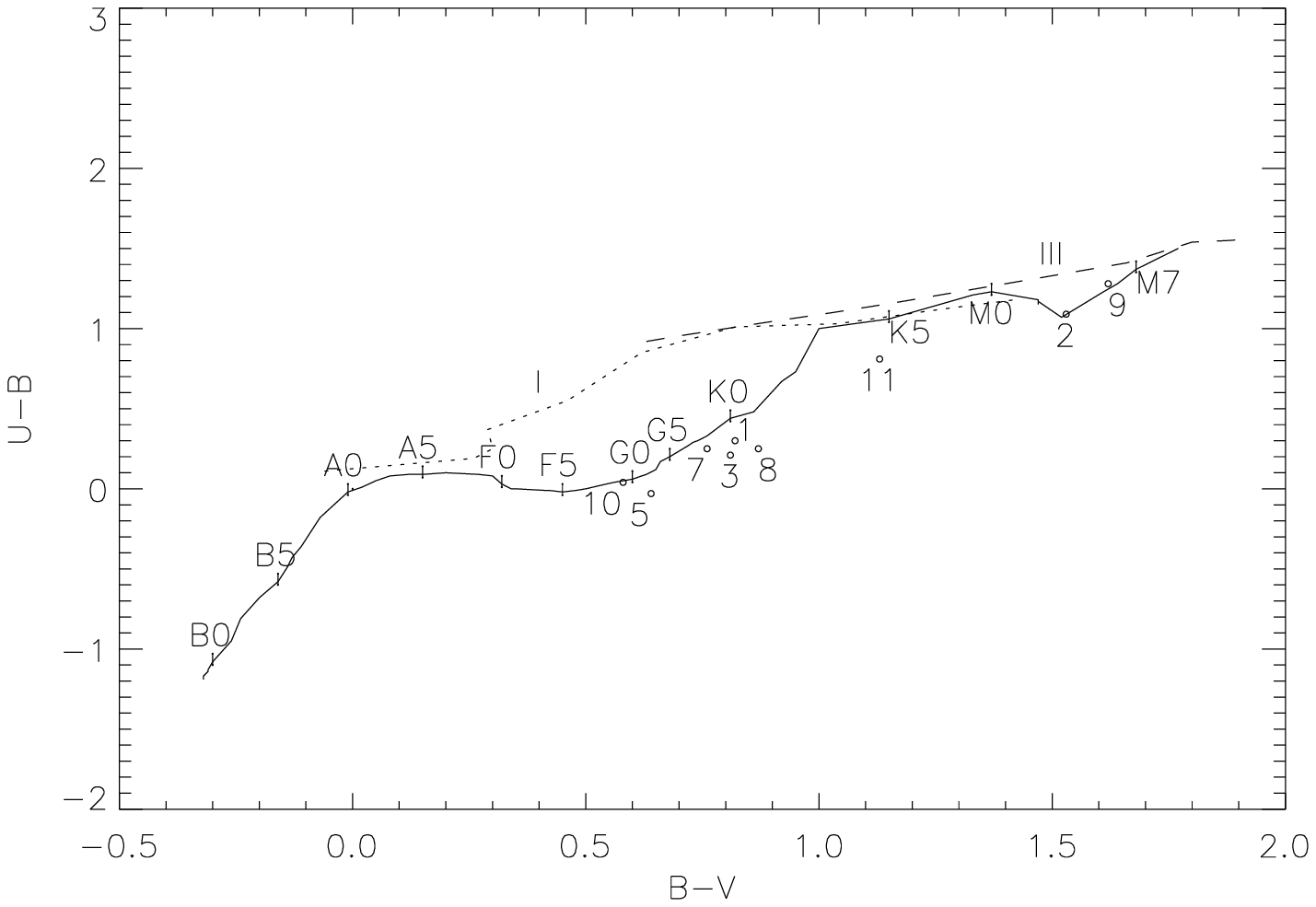}
\caption{The optical B-V versus U-B diagram of the 24$\mu$m excess stars. Each star is symbolized as
small circle and labelled with the number. Stars 4 and 6 are lack, because of problems in some optical
bands.  The dotted line shows the normal dwarf stars in the color-color plot, and the corresponding
spectral types are also labelled.}
\end{figure}

\clearpage
\begin{figure}
\centering
\renewcommand{\thefigure}{4}
\includegraphics[width=\textwidth, angle=0]{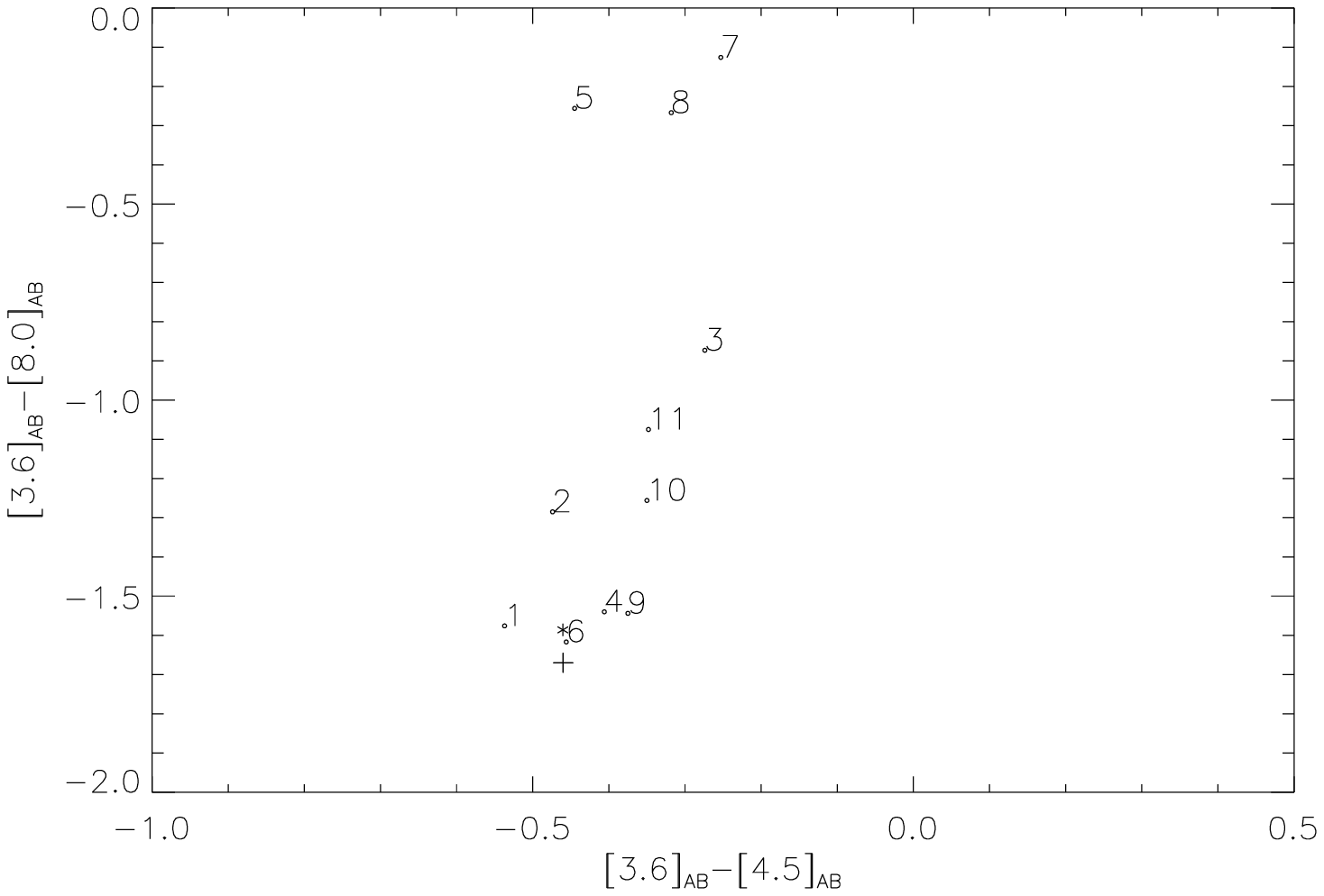}
\caption{ The mid-infrared IRAC color-color diagram of the 24$\mu$m excess stars. All the colors are
in AB magnitude system. Each star is symbolized as small circle and labelled with the number. The star and plus
symbols present the positions of Vega and black-body with temperature of 9600K.  Three stars present a high mid-infrared excess at 8$\mu$m band.
Four present a modest excess and the other four show a minor excess at 8$\mu$m band.
}
\end{figure}

\clearpage
\begin{figure}
\centering
\renewcommand{\thefigure}{5}
\includegraphics[width=\textwidth, angle=0]{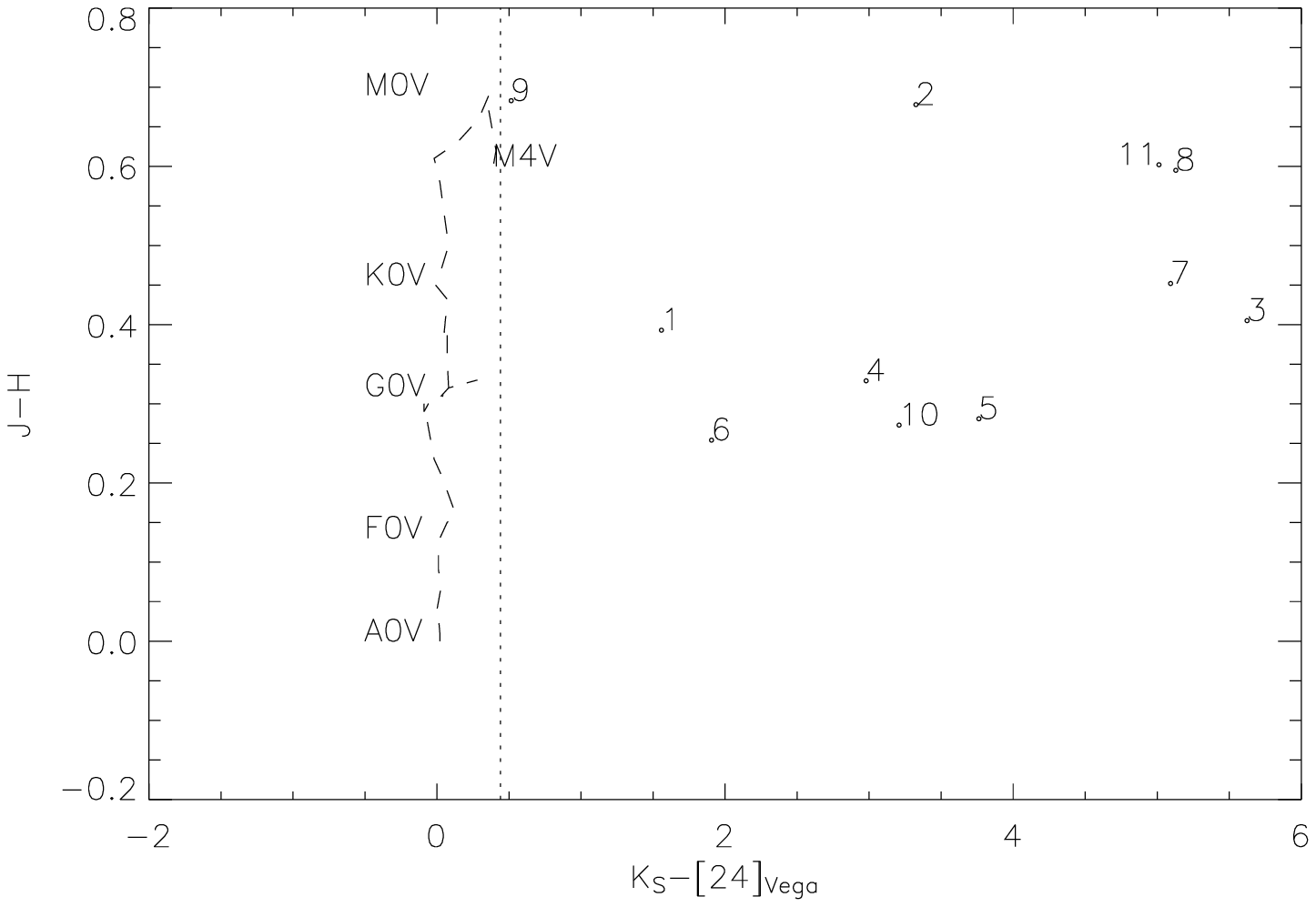}
\caption{The plot of $J$-$H$ versus $K_S$-[24]$_{Vega}$. All the color are in the Vega magnitude system.
Each star is symbolized as small circle and labelled with the number. The dashed curve shows the normal
dwarf stars labelled with the corresponding spectral types. The dotted line gives our criteria to select
the 24$\mu$m excess source. Except Star 9, all the stars have $K_S$-[24]$_{Vega}$ higher than 1 magnitude
and four stars present very strong 24$\mu$m excess.
\label{fig5}}
\end{figure}

\clearpage
\begin{figure}
\centering
\renewcommand{\thefigure}{6}
\includegraphics[width=\textwidth, angle=0]{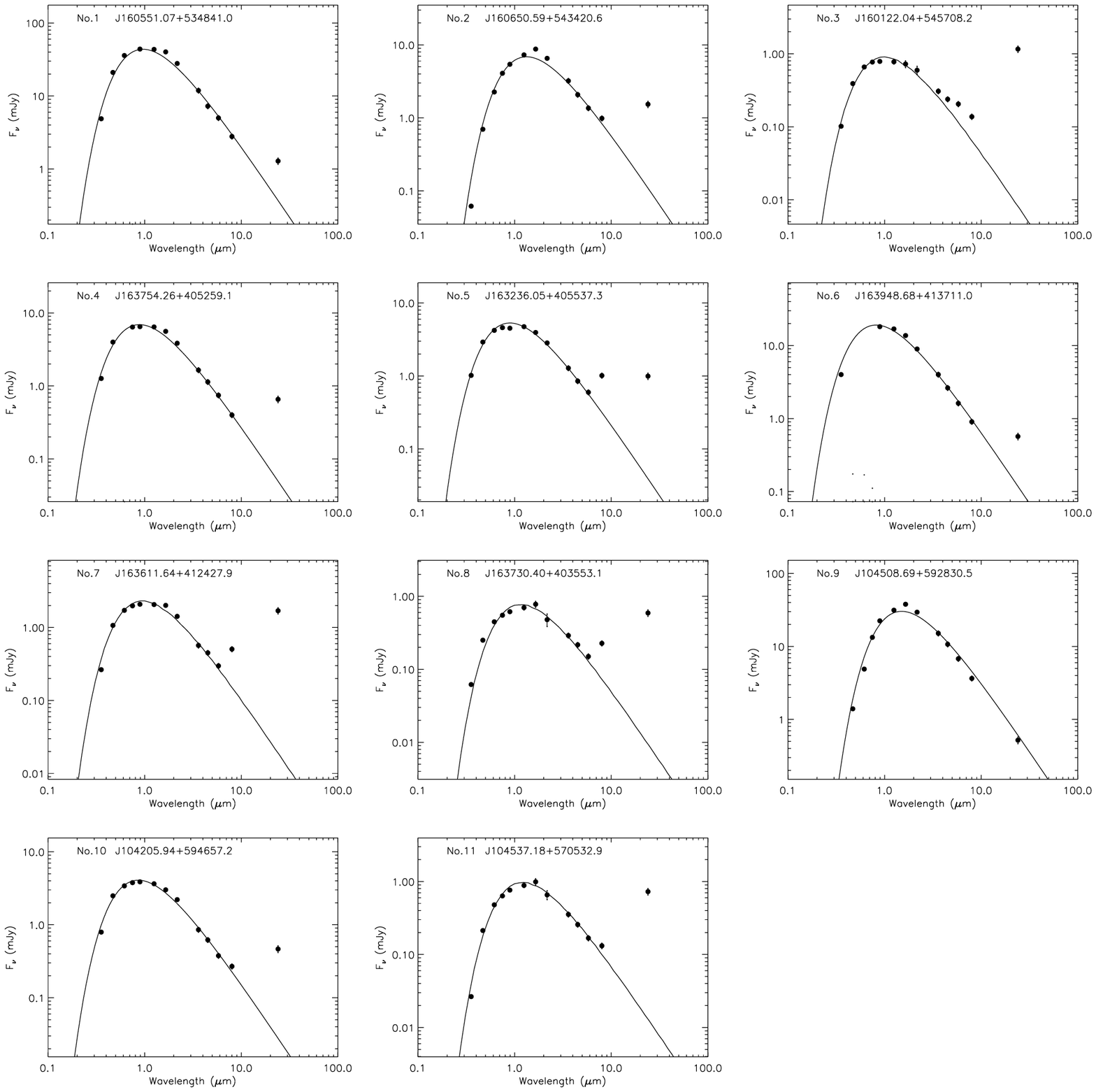}
\caption{The optical to mid-infrared SEDs of the 24$\mu$m excess stars. The black points give the fluxes
of the SDSS five bands ($u$,$g$,$r$,$i$,$z$), the 2MASS three bands ($J$,$H$,$K_S$) and the $Spitzer$
IRAC four bands and MIPS 24$\mu$m band. The photospheric emission is ploted as solid curves. Comparing
         with the solid curves, all these stars
present excess at mid-infrared.
\label{fig6}}
\end{figure}

\clearpage
\begin{sidewaystable}[h]
\vspace{15cm}
\caption[]{The names, magnitudes and colors of the 11 24$\mu$m excess stars.}
\centering
\small
\begin{tabular}{l l rrrrrrrrrrrrrr}
\hline\hline
NO. & Name & $u$ & $g$ & $r$ & $i$ & $z$ & $J$ & $H$ & $K_S$ & [3.6]$_{AB}$ & [4.5]$_{AB}$ & [5.8]$_{AB}$ & [8.0]$_{AB}$ & [24]$_{AB}$ &  $K_S$-[24]$_{Vega}$\\
\hline
1 & J160551.07+534841.0 &14.64 &13.13 &12.52 &13.38 &12.29 &11.41 &11.02 &10.95 &13.71 &14.25 &14.65 &15.29 &16.13 & 1.56\\
  &                 &$\pm$0.02 &$\pm$0.00 &$\pm$0.00 &$\pm$0.00 &$\pm$0.01 &$\pm$0.02 &$\pm$0.02 &$\pm$0.02 &$\pm$0.08 &$\pm$0.08 &$\pm$0.08 &$\pm$0.08 &$\pm$0.10 &$\pm$0.10\\
2 &J160650.59+543420.6 &19.39 &16.83 &15.52 &14.88 &14.56 &13.35 &12.67 &12.52 &15.13 &15.61 &16.07 &16.42 &15.94 & 3.32\\
  &                 &$\pm$0.04 &$\pm$0.02 &$\pm$0.02 &$\pm$0.02 &$\pm$0.02 &$\pm$0.03 &$\pm$0.02 &$\pm$0.02 &$\pm$0.08 &$\pm$0.08 &$\pm$0.08 &$\pm$0.08 &$\pm$0.10 &$\pm$0.10\\
3 & J160122.04+545708.2 &18.84 &17.46 &16.86 &16.69 &16.66 &15.78 &15.38 &15.12 &17.68 &17.96 &18.12 &18.56 &16.24 & 5.62\\
  &                 &$\pm$0.03 &$\pm$0.03 &$\pm$0.02 &$\pm$0.02 &$\pm$0.02 &$\pm$0.07 &$\pm$0.11 &$\pm$0.13 &$\pm$0.08 &$\pm$0.08 &$\pm$0.08 &$\pm$0.09 &$\pm$0.10 &$\pm$0.17\\
4 & J163754.26+405259.1 &16.10 &14.94 &  -   &14.39 &14.37 &13.49 &13.16 &13.10 &15.86 &16.26 &16.72 &17.40 &16.86 & 2.98\\
  &                 &$\pm$0.02 &$\pm$0.01 &$\pm$0.00 &$\pm$0.01 &$\pm$0.02 &$\pm$0.02 &$\pm$0.03 &$\pm$0.03 &$\pm$0.08 &$\pm$0.08 &$\pm$0.08 &$\pm$0.08 &$\pm$0.10 &$\pm$0.11\\
5 & J163236.05+405537.3 &16.34 &15.28 &14.85 &14.76 &14.77 &13.82 &13.54 &13.43 &16.13 &16.58 &16.96 &16.39 &16.41 & 3.76\\
  &                 &$\pm$0.02 &$\pm$0.01 &$\pm$0.01 &$\pm$0.02 &$\pm$0.02 &$\pm$0.03 &$\pm$0.05 &$\pm$0.05 &$\pm$0.08 &$\pm$0.08 &$\pm$0.08 &$\pm$0.08 &$\pm$0.10 &$\pm$0.11\\
6 & J163948.68+413711.0 &14.85 &  -   &  -   &  -   &13.25 &12.44 &12.19 &12.18 &14.90 &15.35 &15.88 &16.51 &17.01 & 1.91\\
  &                 &$\pm$0.01 &$\pm$0.05 &$\pm$0.03 &$\pm$0.03 &$\pm$0.01 &$\pm$0.02 &$\pm$0.02 &$\pm$0.02 &$\pm$0.08 &$\pm$0.08 &$\pm$0.08 &$\pm$0.08 &$\pm$0.10 &$\pm$0.11\\
7 & J163611.64+412427.9 &17.81 &16.37 &15.82 &15.67 &15.60 &14.73 &14.27 &14.18 &17.02 &17.27 &17.72 &17.14 &15.83 & 5.09\\
  &                 &$\pm$0.02 &$\pm$0.02 &$\pm$0.01 &$\pm$0.02 &$\pm$0.02 &$\pm$0.03 &$\pm$0.05 &$\pm$0.06 &$\pm$0.08 &$\pm$0.08 &$\pm$0.08 &$\pm$0.08 &$\pm$0.10 &$\pm$0.12\\
8 & J163730.40+403553.1 &19.38 &17.94 &17.28 &17.06 &16.93 &15.90 &15.30 &15.36 &17.74 &18.06 &18.46 &18.01 &16.97 & 5.13\\
  &                 &$\pm$0.03 &$\pm$0.01 &$\pm$0.01 &$\pm$0.01 &$\pm$0.02 &$\pm$0.08 &$\pm$0.10 &$\pm$0.19 &$\pm$0.08 &$\pm$0.08 &$\pm$0.09 &$\pm$0.08 &$\pm$0.10 &$\pm$0.22\\
9 & J104508.69+592830.5 &18.88 &16.08 &14.69 &13.60 &13.02 &11.77 &11.09 &10.89 &13.45 &13.83 &14.32 &14.99 &17.11 & 0.51\\
  &                 &$\pm$0.05 &$\pm$0.01 &$\pm$0.02 &$\pm$0.02 &$\pm$0.01 &$\pm$0.03 &$\pm$0.03 &$\pm$0.02 &$\pm$0.08 &$\pm$0.08 &$\pm$0.08 &$\pm$0.08 &$\pm$0.11 &$\pm$0.11\\
10&J104205.94+594657.2 &16.61 &15.45 &15.08 &14.97 &14.93 &14.11 &13.83 &13.70 &16.57 &16.92 &17.46 &17.83 &17.23 & 3.21\\
  &                 &$\pm$0.01 &$\pm$0.03 &$\pm$0.02 &$\pm$0.01 &$\pm$0.02 &$\pm$0.03 &$\pm$0.04 &$\pm$0.04 &$\pm$0.08 &$\pm$0.08 &$\pm$0.08 &$\pm$0.08 &$\pm$0.11 &$\pm$0.12\\
11&J104537.18+570532.9 &20.30 &18.12 &17.21 &16.90 &16.69 &15.64 &15.04 &15.02 &17.53 &17.88 &18.34 &18.60 &16.75 & 5.01\\
  &                 &$\pm$0.05 &$\pm$0.02 &$\pm$0.02 &$\pm$0.02 &$\pm$0.02 &$\pm$0.07 &$\pm$0.11 &$\pm$0.14 &$\pm$0.08 &$\pm$0.08 &$\pm$0.09 &$\pm$0.09 &$\pm$0.10 &$\pm$0.18\\
\hline
\end{tabular}
\label{tab:LPer}
\end{sidewaystable}

\clearpage
\begin{table}
\begin{center}
\caption{The log of observation, spectral type and the fractional luminosity
 of the 11 24$\mu$m excess stars.}
\begin{tabular}{llrrrrrr}
\hline\noalign{\smallskip}
NO. & Name & Date of Obs. & Exptime & Instrument & Slit    & Sp. & $f_d$ \\
    &      &              & sec     &            & arcsec  &     &       \\
\hline\noalign{\smallskip}
1 &J160551.07+534841.0 & Feb.  4, 2006 & 900 & OMR 200\AA/mm & 2.5   &G8V &7.0e-5\\
2 &J160650.59+543420.6 & Apr. 28, 2006 &2400 & BFOSC Grism\#4 & 1.8  &M0V &7.4e-4\\
3 &J160122.04+545708.2 & May.  6, 2006 &3600 & OMR 200\AA/mm & 2.0   &K0V &3.2e-3\\
4 &J163754.26+405259.1 & May.  5, 2006 &2700 & OMR 200\AA/mm & 2.0   &G2V &2.1e-4\\
5 &J163236.05+405537.3 & Feb. 21, 2006 &3600 & BFOSC Grism\#4 & 1.1  &G3V &4.2e-4\\
6 &J163948.68+413711.0 & Feb.  4, 2006 &1800 & OMR 200\AA/mm & 2.5   &F8V &5.4e-5\\
7 &J163611.64+412427.9 & May.  5, 2006 &2700 & OMR 200\AA/mm & 2.0   &G7V &1.8e-3\\
8 &J163730.40+403553.1 & May.  6, 2006 &3600 & OMR 200\AA/mm & 2.0   &K4V &2.1e-3\\
9 &J104508.69+592830.5 & Feb.  3, 2006 &2400 & OMR 200\AA/mm & 2.5   &M3V &6.9e-5\\
10&J104205.94+594657.2 & Feb.  3, 2006 &2400 & OMR 200\AA/mm & 2.5   &G0V &2.4e-4\\
11&J104537.18+570532.9 & Feb.  3, 2006 &3600 & OMR 200\AA/mm & 2.5   &K6V &2.3e-3\\
\noalign{\smallskip}\hline
\end{tabular}
\end{center}
\end{table}

\label{lastpage}

\end{document}